\documentclass[10pt]{iopart}
\usepackage{graphicx}
\usepackage{iopams}
\usepackage{bm}

\begin{document}
 


\def\p {\partial}
\def\t {\tilde}
\def\be {\begin{equation}}
\def\ee  {\end{equation}}
\def\bea {\begin{eqnarray}}
\def\eea {\end{eqnarray}}
\def\nn {\nonumber}

\title{Thermodynamics and phases in quantum gravity }

\bibliographystyle{iopart-num}
\author{Viqar Husain}
\address{Department of Mathematics and Statistics, \\
University of New Brunswick, Fredericton, NB E3B 5A3, Canada}
\address{Perimeter Institute for Theoretical Physics,\\
31 Caroline Street North,
Waterloo, ON N2L 2Y5, Canada}
\author{R.B. Mann}
\address{Perimeter Institute for Theoretical Physics,\\
31 Caroline Street North,
Waterloo, ON N2L 2Y5, Canada}

\address{Department of Physics and Astronomy\\
University of Waterloo,
Waterloo, ON N2L 3G1 Canada\\ \mbox{\ \ }}

\begin{abstract}
We give an approach for studying quantum gravity effects on black hole thermodynamics.  This  combines a quantum framework for gravitational collapse with  quasi-local
definitions of energy and surface gravity.  Our arguments suggest that (i) the specific heat of a
black hole becomes positive after a phase transition near the Planck scale,(ii) its entropy
acquires a logarithmic correction, and (iii) the mass loss rate is modified such that Hawking radiation stops  near the Planck scale. These results are due essentially to a realization of fundamental  discreteness in quantum gravity, and are in this sense potentially theory independent.

\end{abstract}

The first law of black hole mechanics  is a result concerning a subspace of the
solution space of Einstein's equations parametrized by  a black hole's mass, charge,
and angular momentum.  Perturbations of these parameters are related, and this
comprises  the first law \cite{carter}.   For the simplest Schwarzschild black hole the
law is
\be
dM = \frac{\kappa}{8\pi}\  {dA}
\ee
Semiclassical arguments due to Bekenstein and Hawking \cite{beck,hawk} have provided a physical and mathematical basis for identification of the surface gravity $\kappa$ with temperature and horizon area $A$ with entropy.

It is interesting to ask how the laws of black hole mechanics might emerge from a theory of quantum gravity.  If such a theory is  a conventional quantum system, one might expect that black holes are described by physical semiclassical states peaked at values of  mass, charge, and angular momentum $(M,Q,J)$.
Such states may be denoted
\be
|M,Q,J\rangle_{sc}(t)
\ee
to indicate that they describe configurations peaked on classical values of the black hole's conserved charges, where $t$ is a parameter characterizing the state's spread.  It is reasonable to conjecture that the theory has operators corresponding to these classical observables, so that one can compute  expectation values such as
\be
\  _{sc}\langle M,Q,J |\hat{M}| M,Q,J \rangle_{sc} = M + O(M, \lambda, t).
\ee
Quantum corrections represented in the second term would in general be present because semiclassical states are not necessarily  eigenstates of  the relevant operators $\hat{M}, \hat{J}$, etc. These corrections would depend on   $t$  and on a  fundamental discreteness scale $\lambda$.  Such states have been constructed  for example for  quantum cosmology  \cite{hw-sc}.

What is required to provide a concrete implementation of these ideas is not just a quantum mechanical model of a static or stationary black hole, but a larger system capable of describing black hole formation and its subsequent evolution. One such system is the gravity-scalar field system in spherical symmetry, which is a  non-trivial $2-$dimensional field theory. It has been well studied classically culminating in the discovery of scaling behavior at the onset of black hole formation \cite{choptuik}. One of the results is that black holes form with arbitrarily small mass for  asymptotically flat initial data that is sufficiently massive. This is the so-called supercritical  data. There is also a class of lower mass  (subcritical) data for which there is no black hole formation. These are  separated by a critical solution which is  a naked singularity. It represents a finely tuned violation of the cosmic censorship hypothesis.

Quantum gravity must modify these results due to at least two considerations: the existence of a fundamental length and  a mechanism of singularity avoidance. Additionally  these features  should  affect the thermodynamical properties of black holes in the Planck regime. This is based on the intuition that any mechanism for singularity avoidance must effectively be a form of  gravitational repulsion. This effectively puts an upper bound  on matter densities, similar to Fermi pressure \cite{hw-ess}. This is turn should  strongly affect the physics of the late stages of Hawking evaporation.

Motivated by these considerations, we study here the effect of a fundamental discreteness scale on the thermodynamic properties of black  holes. We show that the specific heat of a black hole can become positive as the temperature reaches the Planck regime, and that its entropy acquires logarithmic corrections.

Our starting point is a Hamiltonian approach to the problem of quantum gravitational collapse for the same setting that has been well studied for classical collapse, namely Einstein gravity minimally coupled to a massless scalar field. A quantization of this system is  available \cite{hw},  and this is what we use here. An important ingredient  in the quantization is a class of operators that represent the radial function $R(r,t)$ and its inverse in the spatial $3-$metric
\be
ds^2 = q_{ab}dx^a dx^b = dr^2 + R(r,t)^2 d\Omega^2.
\label{met}
\ee
The full configuration space of the model is this radial field together with the scalar field $\phi(r,t)$.

Since the general case is dynamical, it is appropriate to consider thermodynamical variables associated  with evolving horizons whose properties may be captured by gravitational phase space variables.  It is possible to define quasi-local classical variables corresponding to energy and surface gravity that are known to satisfy ``local laws''  of black hole mechanics \cite{hayward}. Using the quantization described in Ref. \cite{hw} it is possible to define operator analogs of these variables as a first step in formulating such laws in quantum gravity.

A suitable definition of quasi-local energy for our purpose is the Misner-Sharp energy  \cite{MS}
\be
E= \frac{1}{2} R (1- q^{ab} \partial_a R \partial_b R).
\ee
Other definitions, such as the Hawking energy, may also be used, but this does not effect the
approach we take in this paper.   This  expression together with the  definition 
\be
\kappa = \frac{1}{2} q^{ab} \p_a\p_b R
\ee
for the surface gravity of a trapping horizon
leads  to the quasi-local first law \cite{hayward}
\be
dE = \frac{\kappa}{8\pi}  dA  - \frac{1}{2} p\ dV.
\label{qloc1}
\ee
Each quantity here is evaluated on the trapping horizon; $A$ is its area,   $V=4/3 \pi R^3$ and
$p = |j^a|$ the norm of the matter current ($j^a = q^{ab}\p_b \phi$ for the scalar field).

For the case where the trapping horizon is slowly varying or not at all, the second term in (\ref{qloc1})
vanishes, and $\kappa$ becomes  inversely proportional to the horizon radius $R_H$. This is familiar from the Schwarzschild solution.  Although our approach applies to the general dynamical case, for simplicity we consider here the situation where the radial size of the local horizon is approximately constant.  Thus  on the horizon we have 
\be
E = \frac{1}{2} R_H,\ \ \ \ \ \ \ \ \ \ \ \ \ \ \ \  \kappa = \frac{1}{R_H}.
\ee
 
With these consideration in mind, we propose the following {\it state dependent} definitions of trapping horizon entropy $S$ and specific heat $C$ entirely within the quantum theory.
\be
 dS = \frac{d\langle \hat{E} \rangle}{ \langle \hat{\kappa} \rangle },\ \ \ \ \ \ \ \ \ \ \ \ \ \
 C = \frac{d\langle \hat{E} \rangle}{ d\langle \hat{\kappa} \rangle},
 \label{S,C}
\ee
where $\hat{E}$ and $\hat{\kappa}$ are suitably defined operators representing the corresponding classical quantities. The differentials in these equations are to be taken  with respect to the  horizon size parameter characterizing the state. This could be the parameter at which a semiclassical 
state is peaked, or an  eigenvalue in case of a state in which the operators are diagonal. We shall use the latter to show  that the familiar entropy formula  modified by logarithmic corrections is recovered,  and that the specific heat exhibits a phase transition from negative to positive values near the  Planck scale.

\begin{figure}
\includegraphics[height=4in,width=5in,angle=0]{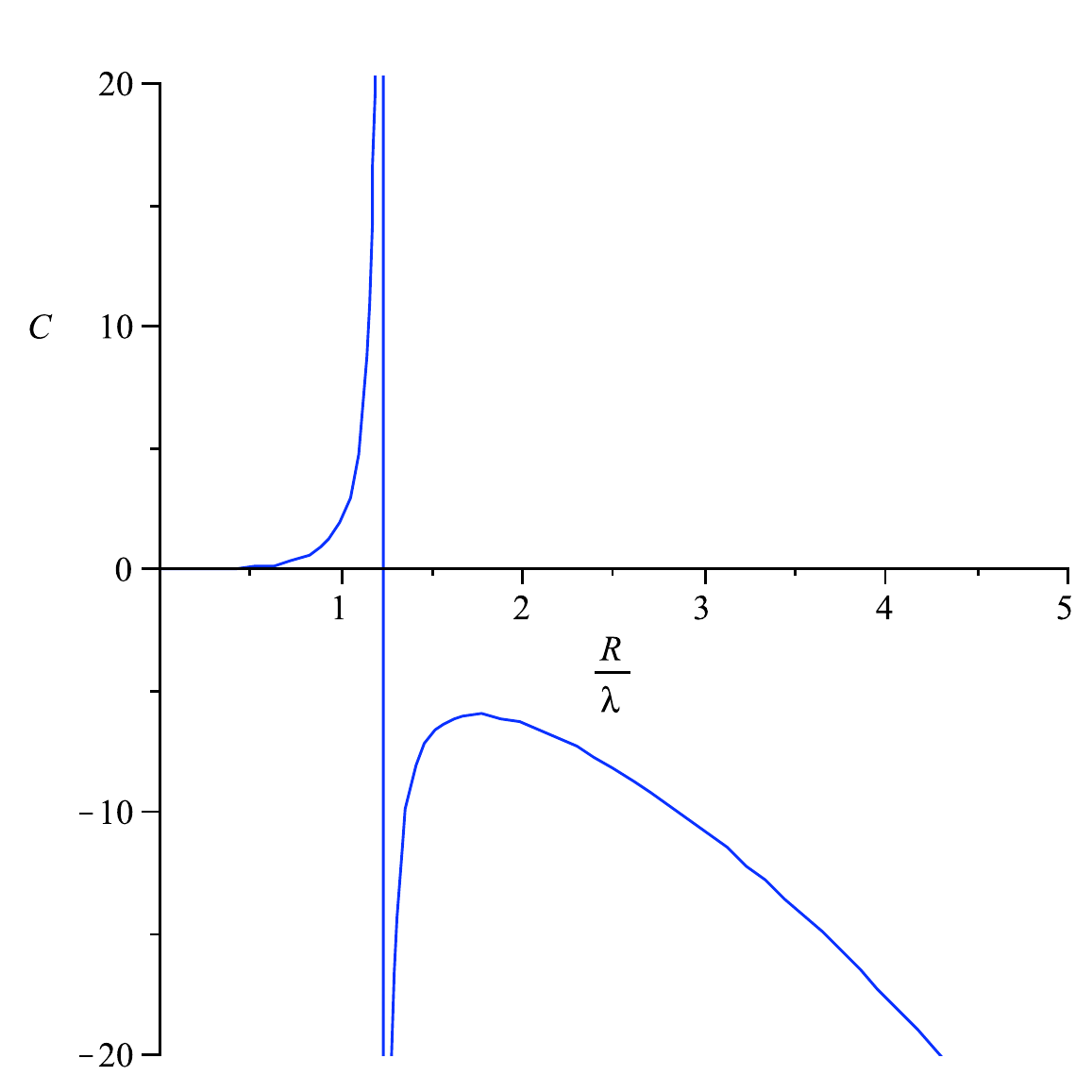}
 \caption{\baselineskip = 1.0em   The specific heat as a function of horizon radius $R/\lambda$ 
 for the case $l=1/5$. The singularity and positive region for small horizons is apparent. For larger horizons the function matches the negative semiclassical result. }
 \label{sheatfig}
\end{figure}

In the quantization described in \cite{hw}, basis states are associated with a radial lattice 
of points   $\{r_k\}$ ($k=1\cdots N$), with field values at the points prescribed. For a theory
with only one field degree of freedom, the basis states are 
\be 
| R_1, R_2, \cdots R_N\rangle.
\ee
For matter coupling, the basis is simply a product of states of this type.
 
The operators representing configuration variables such as $R$ are diagonal on these states  
\be
\hat{R}(r_k) | R_1, R_2, \cdots R_N\rangle = R_k \ | R_1, R_2, \cdots R_N\rangle,
\ee
and inverse configuration operators are realized as
\bea
&& \widehat{\frac{1}{ R(r_k)}_\lambda} |R_1,R_2, \cdots R_N\rangle  \nn\\
&=& \left[   \frac{l}{ 2\lambda}\left( | R_k +\lambda|^{1/l} -  | R_k -\lambda|^{1/l}   \right)\right]^{l/(l-1)}|R_1,R_2, \cdots R_N\rangle \nn \\
&\equiv& f(R_k, \lambda, l)\  |R_1,R_2, \cdots R_N\rangle
\label{1/R}
\eea
 There are two parameters in the definition of the $1/R$ operator. One is the fundamental discreteness scale $\lambda$ and the other is a number $0<l<1$ that characterises an ambiguity in defining this operator. For $r>> \lambda$, the function $f$ behaves like $1/R$, but for sufficiently small $R$ it has $l$ and $\lambda$ dependence.   This is illustrated by the expansion for $R>>\lambda$ :
\be
f(R,\lambda,l)= \frac{1}{R} + \frac{(2l-1)\lambda^2}{6lR^3} + {\cal O}(\lambda^4)
\ee

The entropy and specific heat for a trapping horizon of radius $R_H$ can be computed using equations (\ref{S,C}).  The quantum gravity modifications in these quantities may be illustrated by considering  a lattice basis state $|R_1,R_2, \cdots R_N\rangle$, where the  field value $R_H$ occurs at one of the lattice points. We find
\be
dS = \frac{d\langle \hat{E} \rangle} { \langle \widehat{1/R} \rangle } = \frac{c^2}{2G} \
\frac{R_H\ dR_H}{1+ (2l-1)\lambda^2/6lR_H^2 },
\ee
 \be
S= \frac{c^2}{2G} R_H^2 - \frac{(2l-1)\lambda^2}{12l^2} \ln R_H^2.
\ee
This shows that  the entropy has a $\ln A_H$ correction that depends  on the
discreteness parameter appearing in the inverse $R$ operator, as well as on the $l$ ambiguity in the
definition of this operator.  This correction vanishes for $l=1/2$. Although logarithmic corrections to horizon entropy are known in various approaches \cite{logS}, ours differs in that its origin is intimately tied to the $1/R$ operator, which in turn is connected to singularity avoidance \cite{hw}.

The proposal  (\ref{S,C}) for the specific heat $C$  leads to the quantum gravity modified formula
\be
C = -\frac{2 R_H^4 l}{2 R_H^2 l+ (2l-1)\lambda^2}
\label{qC}
\ee
This formula (\ref{qC}) is derived from an expansion of the function $f(R,\lambda,l)$ in powers of $\lambda$ rather than the full expression.   

For large black holes $R_H>> \lambda$ this gives the familiar negative specific heat, independent of
$\lambda$ and $l$. For $l>1/2$, $C$ is finite and negative for any horizon size. The most dramatic case (illustrated in the figure) occurs for $l<1/2$. The specific heat becomes singular
when a black hole reaches the size
\be
R_{\rm crit}=\lambda\sqrt{\frac{1-2l}{2l}},
\ee
and becomes positive as its size reduces further still. This suggests a second order phase transition in quantum gravity to  a phase where a black hole becomes a thermodynamically stable object.  

If the exact formula for $f(R,\lambda ,l)$ is used rather than its expansion in powers of $\lambda$,
we find that the specific heat has the same qualitative behaviour as in Fig. \ref{sheatfig}, but with the two important differences:  the specific heat  is negative for large $R$ and positive for small $R$ {\it for all 
values of the parameter $l$} (not just for $l<1/2$), and the location of the transition is closer to
$R=0$.  

A further consequence of these arguments is a modification of the black hole mass loss rate. The usual formula may be written as 
\be 
\frac{dM} {dt} = - \alpha\ \frac{M_P}{t_P} \left( \frac{M_P} {M} \right)^2 
\ee 
where $\alpha$ is a positive dimensionless constant and $M_P,t_p$ are the Planck mass and time.   The eigenvalue of the $1/R$ operator (\ref{1/R})  suggests the following modification of the mass loss 
formula:
\be 
\frac{dM}{dt} = - \alpha\ \frac{M_P}{t_P} \left[ M_P \left(  \frac{1}{M} - \frac{(1-2l)\lambda^2}{3l(2M)^3}          \right) \right]^2.
\ee 
It is apparent in this modification that the mass loss rate vanishes for sufficiently small $R$ for $l<1/2$, and that the usual formula is recovered for large $R$. We note again that if the exact formula is used
the rate vanishes at $M=0$ rather than at a value of the order of $M_P$. In either case the evaporation rate slows to zero and a "black hole explosion" is averted. 

Although our arguments are exhibited   using basis states in a polymer quantization of the spherically symmetric sector, it is expected from considerations  of the inverse scale factor in quantum cosmology that qualitatively similar  results will hold for semiclassical states \cite{martin1/a,bt}. Thus these results may reflect a true feature of quantum gravity. 

Our considerations are independent of the value of the Planck scale and therefore apply also to large extra dimension settings. Indeed they depend fundamentally on only two factors: the existence of a fundamental discreteness scale and its effect on the realization of  operators such as $1/R$.  It is in this sense that our arguments are  potentially theory independent. 
The phase transition evident in the specific heat has potentially important  consequences for
black hole production at the Large Hadron Collider   -- it suggests that Hawking evaporation will cease after black holes reduce to  the critical size given by the above formula. Furthermore it provides a quantum gravity basis for  the possibility that missing energies at LHC collisions may be carried  by  Planck scale remnants \cite{lhc}. 

It is of interest to contrast our results and formalism to other works that take a different approach to
the problem of black holes in quantum gravity.  Some early works considered computing black hole temperature  corrections due to back reaction of Hawking radiation on the Schwarzschild geometry.
This is then used to compute the specific heat, which is found to exhibit a divergence and sign change at a critical value of temperature \cite{balbinot-barletta}. This result is qualitatively similar to what we report, although our approach involves non-perturbative quantum gravity considerations, rather than a perturbation theory calculation  of back reaction. 

Other ideas have involved the use of modified uncertainty relations to argue for an upper bound on 
black hole temperature, in some cases motivated by non-commutative geometry \cite{nicolini}. In
these approaches a model of non-commutative gravity is used to extract an analog of a Schwarzschild solution.  This solution is then used to extract thermodynamical properties of black holes is the usual manner. Again some of the qualitative features of specific heat appear similar to what we obtain here 
from canonical quantization of the spherical symmetric sector. These similarities of our results with other
significantly different approaches suggest that some universal considerations involving a fundamental 
 length scale are at play. This is somewhat reminiscent of the entropy-area relation for black holes in that 
whatever postulates are used for microstates, the entropy turns out to be proportional to area\cite{carlip-bhe}.

\medskip

\noindent{\bf Acknowldgements} This work was supported in part by the Natural Science and Engineering Research Council of Canada. The authors thank the Perimeter Institute for Theoretical 
Physics where this work was initiated.

\end{document}